\documentclass[ amsmath,amssymb, aps,reprint, superscriptaddress]{revtex4-1}

\usepackage{graphicx}% Include figure files
\usepackage{dcolumn}% Align table columns on decimal point
\usepackage{bm}% bold math
\usepackage{color}% guess what
\usepackage[normalem]{ulem}
\usepackage{mhchem}

\begin{document}

\preprint{APS/123-QED}

\title{When Quantum Fluctuations Meet Structural Instabilities: The Isotope- and Pressure-Induced Phase Transition in the Quantum Paraelectric NaOH}

\author{Sofiane Schaack}
\affiliation{Sorbonne Université, CNRS UMR 7588, Institut des NanoSciences de Paris, INSP, 75005 Paris, France}
\author{Etienne Mangaud}
\affiliation{Sorbonne Université, CNRS UMR 7588, Institut des NanoSciences de Paris, INSP, 75005 Paris, France}
\affiliation{Univ Gustave Eiffel, Univ Paris Est Creteil, CNRS, UMR 8208, MSME, F-77454 Marne-la-Vallée, France}
\author{Erika Fallacara}
\affiliation{Sorbonne Université, CNRS UMR 7588, Institut des NanoSciences de Paris, INSP, 75005 Paris, France}
\author{Simon Huppert}
\affiliation{Sorbonne Université, CNRS UMR 7588, Institut des NanoSciences de Paris, INSP, 75005 Paris, France}
\author{Philippe Depondt}
\affiliation{Sorbonne Université, CNRS UMR 7588, Institut des NanoSciences de Paris, INSP, 75005 Paris, France}
\author{Fabio Finocchi}
\affiliation{Sorbonne Université, CNRS UMR 7588, Institut des NanoSciences de Paris, INSP, 75005 Paris, France}

% \nodate{}% It is always \today, today,
             %  but any date may be explicitly specified

\begin{abstract} 

 \noindent Anhydrous sodium hydroxide, a common and structurally simple compound, shows spectacular isotope effects: NaOD undergoes a first-order transition, which is absent in NaOH. By combining \textit{ab initio} electronic structure calculations with path integrals, we show that NaOH is an unusual example of a quantum paraelectric: zero-point quantum fluctuations stretch the weak hydrogen bonds (HBs) until they become unstable and break.
 By strengthening HBs via isotope substitution or applied pressure, the system can be driven down to a broken-symmetry antiferroelectric phase.
 We also provide a simple quantitative criterion for HB breaking in layered crystals and show that nuclear quantum effects are crucial in paraelectric to ferroelectric transitions in hydrogen-bonded hydroxides.
\end{abstract}

\maketitle

Beyond the alteration of chemical equilibria and dynamical properties, with important consequences in geology and chemistry, \cite{isotope-effects_book} isotope effects can affect material properties. Changing the nuclear mass in ferroelectrics (FEs) \cite{Horiuchi2019} can actually trigger the mode instabilities in displacive phase transitions \cite{Kvyat2001}.
Understanding these transitions and finely tuning isotope substitution in FEs, which are used in nonvolatile memories and sensors \cite{Lines-Glass}, is an active field of fundamental research with technological applications \cite{Spaldin2019}. Here, we employ molecular dynamics, including quantum effects via path integrals, to explore the ferroelectric behavior of sodium hydroxide, which displays giant isotope effects.

Most studies on FEs focused on perovskites. Generally, the associated energy profile is a double well for the symmetry-broken FE phase, and becomes a single well for the higher-symmetry paraelectric (PE) phase \cite{Chandra_book}. In some cases, the phonon zero-point energy moves below or above the barrier level when the nuclear masses vary. Accordingly, the FE phase can be quenched or restored, as in SrTiO$_3$, a prototypical quantum paraelectric \cite{Muller1979,Marques2005,Bishop2000}. 
More generally, nuclear quantum effects (NQEs) can govern the phase transition, resulting in a critical point that is tuned by variables other than temperature, such as nuclear mass and pressure. Quantum criticality (second-order phase transitions taking place at 0~K) can occur in many systems \cite{Grupp1997,Roussev2003,Chandra2017} 
and motivated the inclusion of quantum fluctuations in Landau-like models \cite{Chandra2017}.

In parallel to bulk and ultrathin perovskite films, \cite{Ghosez2003,Fong2004} other FEs have been investigated, such as hydrogen-bonded crystals \cite{Dalal1998,Koval2002,Ren2018}, whose properties are ruled by the peculiarities of hydrogen bonds (HBs) \cite{Michaelides2011}. Several studies have focused on crystals with strong and short HBs, such as KH$_2$PO$_4$\cite{Koval2002}. In contrast, systems with weak and long HBs were scarcely studied. 
Among them, NaOH is a common material widely used in industrial chemistry \cite{industrial_chem,Yang2015}; a recent study shows evidence of NQEs on the proton transfer rates in water solutions containing NaOH \cite{Hellstrom2018}. However, there is scarce knowledge of the anhydrous NaOH and NaOD crystals, which exhibit several puzzling properties and spectacular isotope effects.

\begin{figure}[]
\includegraphics[width=\linewidth]{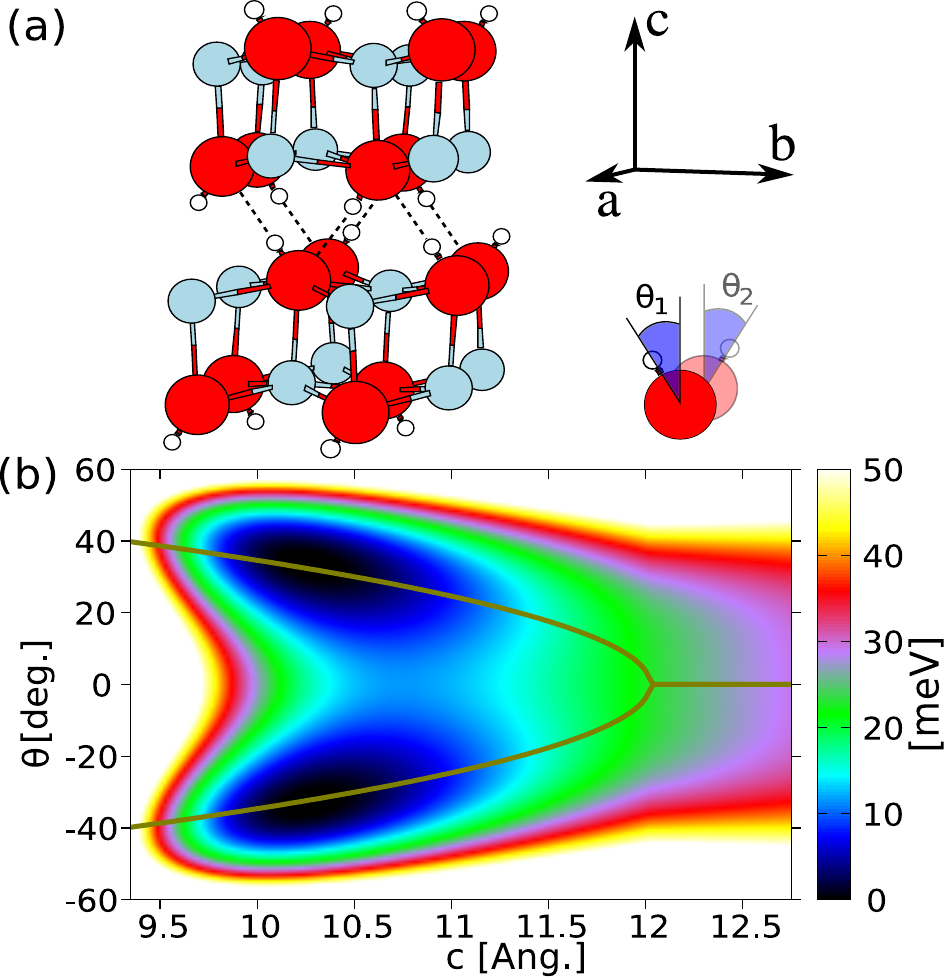}
\caption{\label{fig:NaOH-static} 
The static 0~K picture. 
Panel a: static antiferroelectric (AFE) orthorhombic structure. Na in blue, O in red and H(D) in white. The unit cell is doubled along $\mathbf{b}$, to show the hydrogen bonds as dashed lines. 
The OH groups form angles $\theta_1$ and $\theta_2$ with $\mathbf{c}$ axis.
Panel b: AFE potential energy surface $E(c,\theta)$ versus angle $\theta\!=\!\theta_1\!=\! -\theta_2$ and lattice parameter $c$.  $\displaystyle \theta_0(c)=\min_{\theta} E(c,\theta)$ is plotted in olive green. 
}
\end{figure}

Anomalies in infrared (IR) \cite{Busing1955}, Raman \cite{Kanesaka1982} and NMR spectra \cite{Bastow1986a} have been reported for NaOD, though absent in NaOH.
X-ray, neutron diffraction and dielectric constant measurements reported
a structural change of NaOD around $T_C \simeq 153$K \cite{Bastow1986}. Below $T_C$, NaOD is monoclinic 
with {$P2_1/a$} space group. % (\#14).
Above $T_C$, the structure is orthorhombic with space group $B{mmb}$ 
\cite{Bleif1982}.
Both structures are layered (Fig.\ref{fig:NaOH-static}) with NaOH (NaOD) stacks.
A noticeable effect at the transition is the $\simeq 0.5$~{\AA} expansion of the $c$ axis, normal to the NaOD layers, while lateral parameters are little affected. Both heat capacity \cite{White1986} and low-frequency dielectric constant \cite{Bessonette1999} display anomalies in NaOD around $T_C$, which are absent in NaOH down to 6~K.
Although the H or D distributions were not experimentally determined, a microscopic model where the hydroxyl orientation changes from slanted to vertical was conjectured on the basis of the similarity between the NaOD transition at ambient pressure and 153~K and that in NaOH at 10~kbar and room temperature \cite{Beck1993}. This similarity also raises the question of a quantum phase transition in sodium hydroxide \cite{Chandra2017}. 
Though both transitions were observed experimentally, the microscopic description of the transition mechanism encompassing the giant isotope effects is still lacking and the formation or rupture of interlayer HBs is equivocal.

In order to solve these issues, we investigate the behavior of NaOH and NaOD accounting for NQEs via path-integral based simulations \cite{iPI2014}, using the i-PI package \cite{iPI2019} with the PIGLET thermostat\cite{Ceriotti2011_Piglet, iPI2014}. 
The atomic forces are computed by density functional theory in the Perdew-Burke-Ernzerhof (PBE) approximation \cite{Perdew1996}, using the Quantum Espresso package \cite{QE2009}.
Lattice parameters are optimized via cell relaxation enforcing a target hydrostatic pressure with residual stress tensor components within $\pm 1$~kbar.
Supplemental Material (SM), Sec. A provides further computational details.

\begin{table}[h]
\caption{\label{tab:static} Lattice parameters, OH (ionic-covalent) and O$\dotsm$H (hydrogen) bond lengths in {\AA} as obtained via structural optimization at 0~K in the AFE, FE and PE configurations vs. measured lattice parameters for NaOD \cite{Bastow1986} and NaOH  \cite{Jacobs1985}. $\Delta E$ is the energy increase per unit formula with respect to the AFE configuration. For comparison sake, we reported the halved $a$ AFE lattice parameter.}

\setlength {\tabcolsep} {4.5pt}
\begin{tabular}{lcccccc}
\hline
 & $\Delta E$ (meV) & $a$ &  $b$ & $c$ & OH  & O$\dotsm$H  \\
\hline
AFE & 0 & 3.459 & 3.370 & 10.371 & 0.984 & 2.052 \\
FE & 3.4 & 3.459 & 3.361 & 10.380 & 0.984 & 2.053 \\
PE & 20.4 & 3.387 & 3.374 & 11.481 & 0.975 & 3.545 \\ \hline
&  T (K)& \multicolumn{3}{c}{Experimental \cite{Bastow1986,Jacobs1985}  } \\ \hline
NaOD & 77  & 3.419 & 3.366 & 10.80 & & \\
NaOD & 293 & 3.405 & 3.397 & 11.30 & & \\
NaOH & 147 & 3.389 & 3.383 & 11.334 & & \\
NaOH & 294 & 3.401 & 3.398 & 11.378 & & \\
\hline
\end{tabular}
\end{table}

\noindent{\it Classical picture.}
-- First, we consider several static configurations at 0~K discarding all thermal and quantum nuclear effects (see SM, Sec.~B): paraelectric (PE), ferroelectric (FE) and antiferroelectric (AFE). 
In the PE configuration \cite{Merawa2004}, hydroxyls are parallel to $\textbf{c}$ ($\theta_1\!=\!\theta_2\!=\!0$). In the AFE (FE) configuration (Fig.~\ref{fig:NaOH-static}), hydroxyls at $x\!=\!0$ and $x\!=\!1/2$ are slanted in opposite (parallel) directions along $\textbf{b}$ with angles $\theta_1\!=\!-\theta_2$ ($\theta_1\!=\!\theta_2$). 
The corresponding optimized lattice parameters are reported in Table \ref{tab:static}.
AFE is the lowest-energy configuration. The classical thermal energy $k_BT$ at the experimental transition temperature is $\simeq 12$meV, larger than the AFE/FE energy difference while smaller than the AFE/PE one.
Although its $c$ parameter is close to those measured for NaOD at 293~K and NaOH at all temperatures, the static PE is unstable as confirmed by perturbation theory at the harmonic level: 
the OH bending mode in the (\textbf{bc}) plane ($\theta$ oscillations) has imaginary frequency and drives the PE toward AFE or FE configurations, whose $c$ lattice parameters severely underestimates the experimental ones, more than usual in the PBE approximation \cite{Gillan2016}.

To solve this apparent paradox and apprehend the interplay between hydroxyl angle $\theta$ and lattice parameter {\it c}, we compute the potential energy surface $E(c, \theta)$, shown in Fig.~\ref{fig:NaOH-static} for the AFE (a similar behavior is observed for the FE, see SM, Sec.~C).
The static equilibrium angle $\displaystyle \theta_0 (c)=\min_{\theta} E(c,\theta)$ follows an almost perfect square root dependence $\displaystyle \theta_0 (c)=\pm \sqrt{(\xi-c)/\alpha}$ for $c\le \xi=12.1$~{\AA} and is null for $c \ge \xi$. 
The order parameter $\theta_0(c)$ thus adopts a Landau-like behavior with a transition toward the PE configuration ($\theta=0$) for $c\!\ge\!\xi$. 

A Morse potential nicely fits {the minimum energy at fixed $c$}, $E(c)\equiv E[c,\theta_0(c)]$. As the HB length (O$\dotsm$H) is directly correlated to the $c$ lattice parameter (SM, Sec.~D) $E(c)$ provides a suitable and general criterion to pinpoint the HB breaking in NaOH \footnote{Similar analyses have been done for covalent and metallic bonding in solids: see, e.g. Ref.~\onlinecite{Costescu2014} and references therein.}. 
The critical $c$ value is located at the inflection point of the Morse curve: $c_{cr}\simeq 11.2$~{\AA}, which corresponds to a maximum for the interlayer attractive force at a {threshold (O$\dotsm$H)$_{cr}\simeq 2.4$~{\AA}. Below this value,} the HBs resist to rupture, whereas for (O$\dotsm$H)$>$(O$\dotsm$H)$_{cr}$ they can be considered as broken \cite{Costescu2014}.
Computed elastic constants at 0~K confirm this picture: $C_{33}<0$ for $c\in (11.2 - 11.6) $~{\AA} (SM, Sec.~C).
In this $c$ range HBs, which are the main contribution to the interlayer cohesion, disrupt under stretching and become unable to hold the NaOH layers together. At 0~K, the classical crystal is thus elastically unstable \cite{Mouhat2014} for these $c$ values.

\begin{figure}[!]
\hspace*{-0.06\linewidth}
\vspace*{-0.2 cm}
\includegraphics[width=1.08\linewidth]{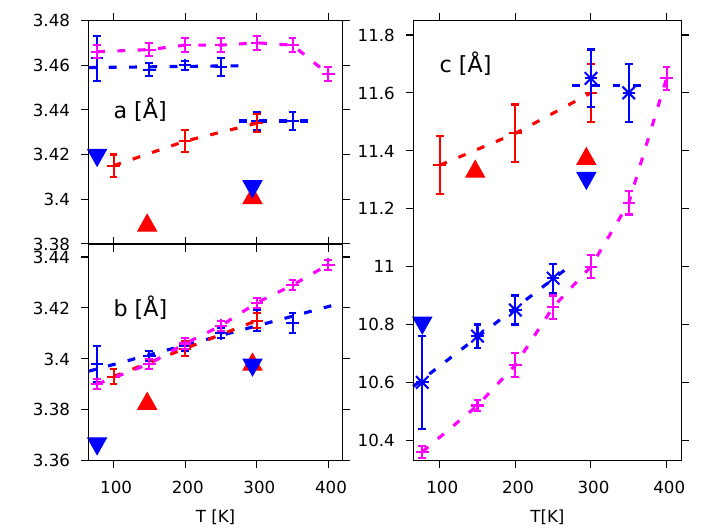}
\caption{\label{fig:abc_vs_T} 
Lattice parameters $a$, $b$, $c$ in~{\AA} vs $T$ as given by PIMD for NaOH (red), NaOD (blue) and classical Langevin dynamics (magenta). The dashed lines are a guide for the eye. Experimental values from Ref.~\onlinecite{Bastow1986} for NaOD (blue triangles) and Ref.~\onlinecite{Jacobs1985} for NaOH (red triangles).}
\end{figure}

In order to appraise the role of \emph{classical} thermal fluctuations, we optimized the lattice parameters between 77 and 400~K (Fig.\ref{fig:abc_vs_T}) through several $(NVT)$ runs. 
At low temperature, the HBs display small fluctuations around the linear configuration {and the lattice parameters are very close to those obtained at 0~K}. When raising $T$, the HBs break and reform, as revealed by the analysis of O-H$\dotsm$O angles and distances.
HBs stretch with $T$ and the crystal %dilates
expands along ${c}$ with a huge coefficient of thermal expansion $\alpha_c \simeq 2.5 \times 10^{-4}$~K$^{-1}$, that further increases beyond 350~K; this huge $\alpha_c$ is the signature of weak interlayer HBs \footnote{We note that $\alpha_c$ is one or 2 orders of magnitude larger than $\alpha_a$ and $\alpha_b$.}.
The classical picture thus reveals the onset of HB breaking. However, several major discrepancies with the experiments still hold: first, $c(T)$ {remains underestimated and} evolves with no abrupt changes, though some anomalies can be detected beyond 350~K (see SM, Sec.~D); second, classical simulations cannot distinguish between NaOD and NaOH and are thus unable to describe the isotope-specific transition.
We discuss below how nuclear quantum effects (NQEs) reshape the classical picture and reconcile theory with experiments.

\noindent{\it Quantum picture.}
-- The inclusion of NQEs via path-integral molecular dynamics (PIMD) changes the classical scenario drastically. First, we optimize the lattice parameters through the same procedure as in classical MD (SM, Sec.~E). 
As HBs are weak and elongated, lateral quantum fluctuations of H and D nuclei dominate over longitudinal ones and further destabilize the HBs \cite{Benoit2005,Michaelides2011}. This yields a much larger lattice parameter $c$ than the classical one, particularly for the lighter isotope: $c$(NaOH)=11.35~{\AA} at 100~K, with $\alpha_c \simeq 10^{-4}$~K$^{-1}$ between 100 and 300~K (Fig.\ref{fig:abc_vs_T}).
In NaOD, $c$ increases linearly from 10.60~{\AA} (77~K) to 10.96~{\AA} (250~K) and abruptly jumps to $\simeq 11.6$~{\AA} (300~K). At the same $T$, the $a$ parameter drops by 
$\simeq 0.02$~{\AA}, consistently with the experiments \cite{Bastow1986}.
This correlated behavior is congruous to a first-order transition between 250 and 300 K in NaOD, though the transition temperature $T_C$ is overestimated with respect to the experimental 153~K value \cite{Bastow1986}. 
Noteworthy, the PIMD simulations faithfully reproduce the giant structural isotope effect and confirm the jumps in the $c$ and $a$ lattice parameters that could be hypothesized but not firmly asserted from the available experimental data.

\begin{figure}[h!]
\includegraphics[width=0.8\linewidth]{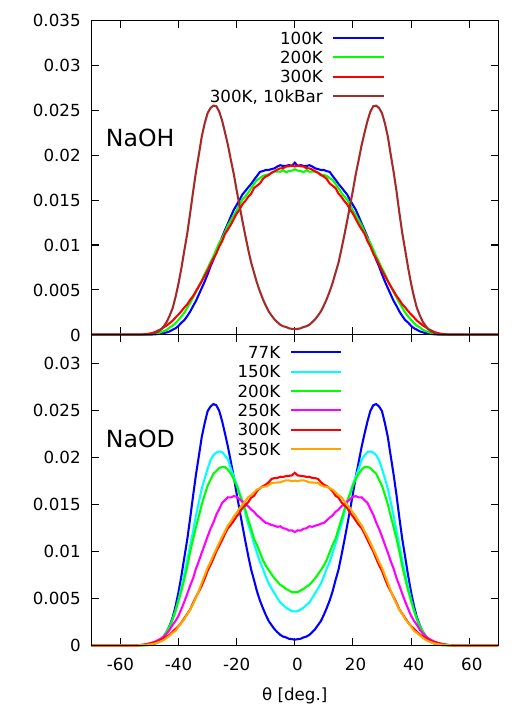}
\vspace*{-0.4cm}
\caption{PIMD Distributions $P(\theta)$ at various temperatures in NaOH and NaOD at ambient pressure; the NaOH case at $T=300$~K under 10 kbar pressure is also shown. } 
\label{fig:p-theta}
\end{figure}
The disruption of the interlayer hydrogen bonds modifies also drastically the orientation of the OD groups: from bimodal distributions with maxima at $\pm\theta_0$ ($\simeq 30^\circ $) for $T<250$~K, the OD angular distributions display a single maximum at $0^\circ$ for $T\ge 300$~K (Fig.\ref{fig:p-theta}).
The structural transition in NaOD is therefore linked to a deep change of the dipole ordering; 
the distribution of the order parameter $\theta$ follows a universal Landau-like behavior as also found in the centering of the hydrogen bonds in KH$_2$PO$_4$ \cite{Koval2002} and ice under extreme pressures \cite{Bronstein2014}.
The absence of transitions in NaOH is confirmed by the persistence of a unimodal OH angular distribution at all temperatures here probed, with a flat profile at low $T$, typical of quantum paraelectrics \cite{Stoll1976}.

\begin{figure}[h!]
\hspace*{-5mm} 
\includegraphics[width=1.05\linewidth]{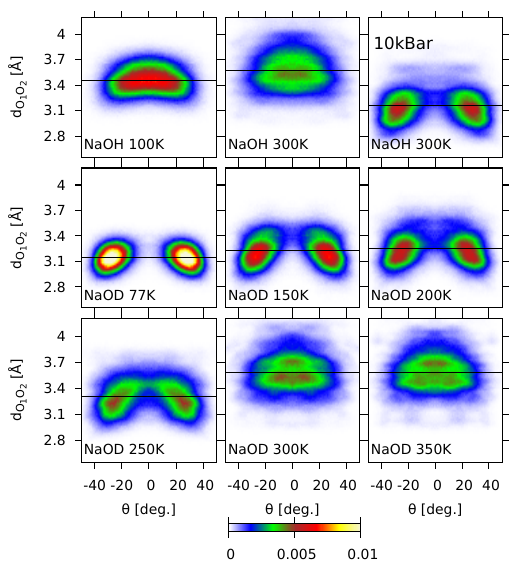}
\vspace{-0.5cm}
\caption{Joint interlayer O--O distance and OD (OH) polar angle probability distributions $P(\theta,d_\textrm{OO})$ in NaOD (NaOH) at various temperatures and ambient pressure. 
Top right panel: NaOH at 300~K under 10~Kbar hydrostatic pressure.
The horizontal lines denote the mean distance $\langle d_\textrm{OO} \rangle$.  
} \label{fig:NaOD_OO-theta}
\end{figure}

The $c$ abrupt change in NaOD correlates with a 5\% discontinuity of the interlayer O--O distance $d_\textrm{OO}$, due to the combined $c$ expansion and in-plane relaxation of the O anions, concomitant to the HB weakening (SM, Sec.~D).
The joint probability $P(\theta,d_\textrm{OO})$ shows a strong correlation between $d_\textrm{OO}$ and the OH polar angle $\theta$ (Fig.\ref{fig:NaOD_OO-theta}). In NaOD at 77~K, $\langle d_\textrm{OO} \rangle = 3.14$~{\AA}, while $\langle d_\textrm{OO} \rangle \ge 3.57$~{\AA} at $T\ge 300$~K, in a dynamical PE state where $\theta$ oscillates around 0$^\circ$.
In contrast, NaOH shows a single probability maximum at $\theta=0^\circ$, with $\langle d_\textrm{OO} \rangle \ge 3.45$~{\AA} at all temperatures.

\begin{figure}
\hspace*{-0.025\linewidth}
\includegraphics[width=1.1\linewidth]{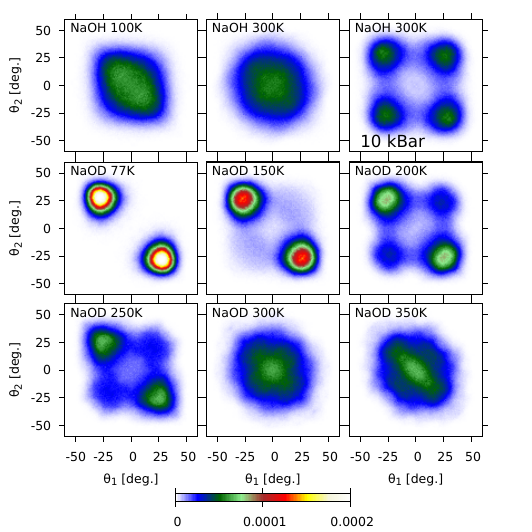}
\vspace{-0.8cm}
\caption{\label{fig:theta-theta} Joint probability distribution of the polar angles of hydroxyl groups at $x\!=\!0$ ($\theta_{1}$) and $x\!=\!1/2$ ($\theta_{2}$) in  NaOH (top) and NaOD (center and bottom) at various temperatures and ambient pressure.
Top right panel: NaOH at 300~K under 10~kbar hydrostatic pressure.}
\end{figure}

The joint probability $P(\theta_1,\theta_{2})$ (Fig.\ref{fig:theta-theta}) of the polar angles of parallel HB chains at $x\!=\!0$ and $x\!=\!1/2$ crystal coordinates (Fig.\ref{fig:NaOH-static}) reveals angular correlations, with remarkable differences between NaOH and NaOD. 
At $77$~K, NaOD is AFE. For increasing $T$, the joint probability also displays tiny secondary maxima corresponding to FE configurations, consistently with the 3.4 meV classical energy difference between the two configurations (Table~\ref{tab:static}). 
The PE configuration is always a local minimum for $T\le 250$~K. 
At 250~K, the angular correlations in NaOD become strongly blurred and at 300~K the two maxima merge into a single one at $\theta_1\!=\!\theta_2\!=\!0$ (the paraelectric state). 
In contrast, the NaOH distribution shows this single maximum at all temperatures and a slightly more elliptic distribution at low $T$ with the main axis along the $\theta_1\!=\!-\theta_2$ AFE line. This is characteristic of a quantum PE \cite{Stoll1976} in which the AFE ordering is prevented by zero-point fluctuations (larger for H than for D).

\noindent{\it NaOH under pressure.} -- The phase transition can be recovered in NaOH by an external pressure that brings the layers (and the facing O anions) closer together and potentially restores hydrogen bonds.
The experimental compressibility is anisotropic, with an abrupt $\sim7\%$ volume contraction at $P_c=10$~kbar upon increasing pressure \cite{Beck1993}.
We simulated NaOH under a 10~Kbar hydrostatic pressure, which mainly impacts the lattice parameter $c$ that recovers its computed value for NaOD at 77~K and ambient $P$ (Table~\ref{tab:pressure}).
The computed volume contraction of NaOH between 0 and 10 kbar amounts to 9\%, close to the 12\% as measured at 297K \cite{Beck1993}. 

\begin{table}[h]
\caption{\label{tab:pressure} Computed lattice parameters for NaOH at $P=10\pm 1$ kbar and 300~K and NaOD at $P\!=\!0\pm 1$ kbar and 77~K.}
 \setlength {\tabcolsep} {15pt}
\begin{tabular}{lccc}
\hline
 & $a$~({\rm\AA}) &  $b$~({\rm\AA}) & $c$~({\rm\AA}) \\
\hline
NaOH  & 3.449 & 3.401 & 10.59 \\
NaOD  & 3.463 & 3.398 & 10.60 \\
\hline
\end{tabular}
\end{table}

The atomic distributions show that 
the contraction of the interlayer distance under pressure actually reinforces the HBs in NaOH and displaces its ground state toward a combination of AFE and FE configurations.
The joint probability $P(\theta_1,\theta_{2})$ (Fig.\ref{fig:theta-theta}) indeed displays four comparable maxima around $\theta_1\!=\!-\theta_2\!=\!\pm 30^\circ$ (AFE) and $\theta_1\!=\!\theta_2\!=\!\pm 30^\circ$ (FE). 
The HBs in NaOH under pressure and room temperature are restored; however, the polar angle ordering contrasts with that in NaOD at low $T$. Indeed, at room temperature, the dipole-dipole interaction that would favor the AFE ordering is negligible with respect to the thermal activation energy \footnote{The energy difference between the {\em static} AFE and FE configurations roughly corresponds to 40~K (Table~\ref{tab:static}).}.
In passing, we note that pressure induces the symmetry-breaking transition in NaOH, while it has opposite effect in KH$_2$PO$_4$ \cite{Koval2002} and in ice VII \cite{Benoit2005,Bronstein2014}. This apparent contradiction can be explained by the nature of HBs \cite{Michaelides2011}, which are very weak in NaOH but strong in both KH$_2$PO$_4$ and ice VII.
Our findings corroborate the observed phase transition in pressure \cite{Beck1993} and suggest that NaOH is close to a quantum critical point for the AFE-FE-PE transitions, in which pressure has partly analogous effects to the H $\rightarrow$ D substitution.
Further studies, beyond the scope of this paper, are necessary to locate the critical point \cite{Chandra_book,Chandra2017}.

\noindent{\it Conclusions.} -- Sodium hydroxide displays a complex behavior that stems from very weak interlayer hydrogen bonds (HBs). 
Our analysis of the classical potential energy surface provides a precise criterion for the transition from elastically stable to unstable HBs under stretching, and locates at the inflection point the critical O--O distance. 
This criterion can be generalized to other layered hydroxides, such as LiOH and KOH \cite{Pagliai2006,Fallacara2021}.
In sodium hydroxyde, interlayer O--O distances are close to their critical value. Quantum fluctuations drive NaOH into the elastically unstable range, where HBs are unable to ensure the interlayer cohesion, in sharp contrast with the classical picture. The order parameter OH polar angle distribution displays a single maximum at all temperatures here probed, and NaOH can be characterized as a quantum paraelectric.
Deuteration reduces quantum fluctuations and restores weak HBs in NaOD, which is antiferroelectric at low $T$.
Further thermal fluctuations besides the quantum ones destabilize HBs: our simulations indicate that NaOD undergoes a first-order phase transition in temperature, as marked by a jump in the interlayer spacing. 

We emphasize that the isotope- and pressure-driven phase transitions in sodium hydroxide escape classical statistical mechanics. Neither the proton nor the deuterium behaviors can be captured by harmonic zero-point corrections on top of a classical picture: the potential energy surface is highly anharmonic and sensitive to mode coupling.
The consistent treatment of the intrinsically quantum difference between H and D via path-integrals instead provides a comprehensive picture of isotope and pressure effects. Our study suggests that NaOH is close to a quantum critical point \cite{Chandra2017} and opens perspectives for tuning the transition in sodium hydroxide by doping, isotope substitution, pressure and coupling to other, more conventional, ferroelectrics. 
In addition, because of its structural simplicity, anhydrous NaOH is also a prototype to study the relation between hydrogen bonding and mechanical properties, which is crucial in supramolecular materials, such as hand-twistable crystals \cite{Desiraju2018}, aza-heterocycles with wavelike topology \cite{Shivakumar2012} and self-healing tricarboxylic acid \cite{Shi2018}.

\begin{acknowledgements}
We thank Sara Bonella for fruitful discussions and for her critical reading of the manuscript, and the referees for constructive criticisms. This work was granted access to the HPC resources of CINES under the allocation A0050906719 and of TGCC under the allocation A0130906719, made by GENCI.
\end{acknowledgements}

 \bibliography{biblio}

\end{document}